\documentclass[aps,pre,reprint]{revtex4-2}

\usepackage{amsmath,amssymb,amsfonts,bm,mathtools}
\usepackage{graphicx}
\usepackage{hyperref}
\usepackage{xcolor}
\usepackage{physics}
\usepackage{booktabs}
\usepackage{multirow}
\usepackage{array}

\hypersetup{
  colorlinks=true,
  citecolor=blue!50!black,
  linkcolor=blue!50!black,
  urlcolor=blue!50!black
}

\newcommand{\ReL}{\operatorname{Re}}
\newcommand{\diag}{\operatorname{diag}}
\newcommand{\vect}[1]{\bm{#1}}

\begin{document}

\title{Functional Dismantling of Network Relaxation through Slow-Branch Susceptibility}

\author{Kaiming Luo}
\email{kmluo24@m.fudan.edu.cn}
\affiliation{College of Future Information Technology, Fudan University, Shanghai 200438, China}

\author{Huiying Zhou}
\affiliation{College of Science and Technology, University of Bordeaux, 33405 Talence, France}
\affiliation{College of Law, Political Science, Economics and Management, University of Bordeaux, 33608 Pessac, France }
\date{\today}

\begin{abstract}
Robustness of relaxation on asymmetric networks is not determined by connectivity alone, because the slow collective mode can be complex and may change its spectral identity under adaptive damage. 
We introduce a slow-branch susceptibility framework for functional dismantling of network relaxation. 
Starting from the projected relaxation dynamics, we show that the relevant robustness observable is the real part of the selected nonzero Laplacian branch, which controls the long-time decay of the nonstationary sector. 
Node deletion is then treated as a dimension-changing compression of the operator, leading to a modal susceptibility (MS) score that estimates the first-order reduction of the branch-tracked relaxation rate from the biorthogonal support of the slow mode. 
In the reciprocal limit, the same construction reduces to the weighted Fiedler sector, placing directed and weighted-undirected networks within a common spectral-response formulation. 
Tests on synthetic and real-world networks show that MS identifies vulnerability patterns that differ from standard centrality-based attacks and edge-level spectral proxies. 
These results resolve a modal-selection ambiguity in non-Hermitian robustness analysis and provide a spectral basis for functional dismantling in asymmetric networks.
\end{abstract}

\maketitle

\section{Introduction}

Asymmetric interactions are a generic feature of nonequilibrium networked systems rather than a marginal complication. Transport in infrastructure and flow networks, influence propagation in biological and social systems, current redistribution in power grids, and synchronization mediated by effectively unidirectional couplings all generate coarse-grained operators that are not reciprocal \cite{Newman2010Networks,AlbertBarabasi2002,RMPArenas2008,RestrepoOttHunt2006,Chung2005Laplacians}.  Recent studies have further shown that collective dynamics in complex systems can be strongly reshaped by directed couplings, and non-normal response channels~\cite{Battiston2021HigherOrderPhysics,Gallo2022DirectedHigherOrder,Zhang2023HigherOrderCollectiveDynamics,Nazerian2024Syncreactivity}. In such systems, robustness is not only a structural question of whether nodes remain connected, but also a functional question of whether the damaged network can still support the collective process carried by the original system.

This distinction is central to many physical and biological networks. A brain network may preserve its anatomical or structural backbone while losing the functional mode associated with coherent activity or information integration. A power grid may remain topologically connected while its slow recovery channel, load-redistribution mode, or synchronization stability is strongly degraded. More generally, a graph can survive as a combinatorial object while the dynamical mode responsible for relaxation, redistribution, or consensus is weakened or reorganized \cite{Forrow2018,Bronski2014,Shahrivar2017}. Functional dismantling therefore differs from structural dismantling: the relevant failure is not necessarily disconnection, but the loss of the spectral channel that sustains the process. This distinction is consistent with recent views of network robustness in which degradation is assessed not only through connectivity loss, but also through the collapse of dynamical function, collective response, or system-level performance~\cite{Grassia2021MachineDismantling,Artime2021FeaturePercolation,Nazerian2024Syncreactivity}.

Spectral viewpoints provide a natural way to formulate this idea. In reciprocal networks, the slow nontrivial Laplacian mode links structure and dynamics through the Fiedler sector \cite{Fiedler1973,Merris1994Laplacian,Mohar1991Laplacian}. Once reciprocity is broken, however, the situation becomes more delicate. The relevant branch can be complex, left and right modal supports separate, and nearby branches can exchange their physical roles under perturbation. The real part of a slow branch controls asymptotic decay, whereas the imaginary part describes rotational or oscillatory modulation. Thus, robustness in directed networks requires identifying which spectral branch carries the long-time functional response.

Adaptive node removal exposes this issue sharply. Classical dismantling strategies rank nodes by degree, betweenness, efficiency loss, reachability, or related centralities \cite{AlbertJeongBarabasi2000,Cohen2001Breakdown,HolmeKim2002Attack,CrucittiLatoraMarchiori2004,Schneider2011,Gao2012,Buldyrev2010}. These measures are effective for structural degradation, but they do not necessarily identify the node whose removal most strongly perturbs the slow dynamical branch. Spectral attack methods improve this connection by using eigenmode information, but edge-based perturbation formulas cannot be directly transferred to node deletion because removing a node deletes a coordinate, all incident edges, and the corresponding normalization of the operator\cite{Luo2026FGIA,Jiang2024FiedlerContribution,Zhou2024ConnectivityRankIndex,Hu2025AcceleratingSynchronization,Jiang2023KeyCycles,wen2025dynamics}. Node attack is therefore a dimension-changing spectral compression problem rather than a standard fixed-dimensional perturbation.

A second ambiguity concerns modal selection during damage. If eigenvalues are independently re-sorted after each deletion, a genuine weakening of the functional decay channel can be confused with a change of spectral bookkeeping. If the evaluation stops at the first loss of strong connectivity, the remaining nonzero relaxation response is discarded even though it may still control the projected dynamics. A useful robustness framework should therefore track the decay observable itself and continue to evaluate the surviving nonstationary sector beyond purely combinatorial thresholds.

In this work, we formulate a branch-aware framework for functional robustness in directed, undirected, and weighted-undirected networks. Starting from the projected relaxation dynamics, we show that the relevant scalar metric is the real part of the slowest nonzero Laplacian branch. We then derive a modal-susceptibility (MS) score that estimates the first-order compressed response of this branch under node deletion. The score ranks nodes by their biorthogonal support on the slow mode and by the predicted reduction of the dominant decay channel.

This construction shifts the attack objective from structural prominence to functional impact. A high-MS node need not be a generic hub; depending on the network, it may appear as a bridge, a shortcut junction, or a distributed relay carrying the slow branch. In the symmetric limit, the same theory reduces to a node-level compression of the weighted Fiedler sector, placing directed and reciprocal networks within a common spectral-response formulation. Across canonical and real-world networks, the resulting rankings reveal vulnerability patterns that differ from centrality-based attacks and structural dismantling, supporting the view that relaxation robustness is organized primarily by the sensitivity of the slow spectral branch.

\section{Relaxation metric and slow-branch selection}
\label{sec:framework}

This section derives the robustness observable from the underlying relaxation dynamics. The point is not to declare $\ReL(\lambda_2)$ by analogy. It is to show that once the decaying sector of the dynamics is defined properly, the natural scalar metric is exactly the real part of the slowest nonzero branch. Establishing that fact first is essential, because the attack algorithm in Sec.~\ref{sec:algorithms} is built to perturb this specific observable rather than a generic graph property.

Consider a finite networked system with $N=\abs{V}$ degrees of
freedom. Its state is represented by the node-state vector
\(
\vect{z}(t)=\left(z_1(t),\ldots,z_N(t)\right)^T ,
\)
where $z_i(t)$ denotes the scalar state of node $i$. The asymmetric
weighted adjacency matrix is denoted by $A=(w_{ij})$, and the
corresponding Laplacian-type generator is
\begin{equation}
L = D-A,\qquad
D=\diag\!\left(d_1,\ldots,d_N\right).
\label{eq:directed-laplacian}
\end{equation}
Throughout the paper we use the row-Laplacian convention: $w_{ij}$
denotes the weight of the directed interaction from node $i$ to node
$j$, and
\(
d_i=\sum_j w_{ij}.
\)
Equivalently, the action of $L$ on a state vector is
\(
(L\vect{z})_i
=
\sum_j w_{ij}\left(z_i-z_j\right).
\)

The linear relaxation dynamics generated by $L$ is then
\begin{equation}
\dot{\vect{z}}(t)=-L\vect{z}(t),
\qquad
\vect{z}(t)=e^{-Lt}\vect{z}(0).
\label{eq:relaxation-dynamics}
\end{equation}
With this convention, the directed weight $w_{ij}$ contributes to the
diagonal entry of row $i$ and to the off-diagonal coupling between
nodes $i$ and $j$. This convention is used consistently in the
node-deletion formulas below.

We assume throughout that the edge weights are nonnegative and that
$L$ is a Laplacian-type generator whose nonzero spectrum lies in the
closed right half-plane for the graph states considered. Zero
eigenvalues are allowed and correspond to stationary or marginal
sectors. The robustness metric below is therefore defined on the
projected decaying sector rather than on the full state space. This
convention is essential for directed graphs, where loss of strong
connectivity may increase the dimension of the zero eigenspace without
eliminating all nonzero relaxation channels.

Such generators arise naturally in diffusion-like transport,
asymmetric consensus, current redistribution, and linearized
synchronization on nonreciprocal substrates
\cite{RMPArenas2008,RestrepoOttHunt2006,Chung2005Laplacians}.

Let $\Pi_0$ denote the spectral projector onto the zero eigenspace of
$L$, and let
\begin{equation}
Q_0=I-\Pi_0
\label{eq:decaying-projector}
\end{equation}
be the complementary projector onto the decaying sector. The
nonstationary component of the dynamics is therefore
\begin{equation}
\vect{u}(t)
=
Q_0\vect{z}(t)
=
Q_0 e^{-Lt}Q_0 \vect{u}(0).
\label{eq:decaying-dynamics}
\end{equation}
Stationary or marginal zero modes should not be counted as relaxation,
so the graph-state robustness is defined by the asymptotic decay
exponent of the restricted semigroup:
\begin{equation}
\mathcal{M}(G)
=
-\limsup_{t\to\infty}
\frac{1}{t}
\log
\norm{Q_0 e^{-Lt}Q_0}.
\label{eq:metric-definition}
\end{equation}
Equation~\eqref{eq:metric-definition} is the starting definition used
in this paper. It depends only on the long-time decay of the
nonstationary sector and is therefore insensitive to purely
combinatorial bookkeeping. The projector in
Eq.~\eqref{eq:decaying-projector} and the reduced dynamics in
Eq.~\eqref{eq:decaying-dynamics} make explicit that the metric is
defined on the decaying sector alone.

To evaluate Eq.~\eqref{eq:metric-definition}, write the nonzero spectral decomposition of $L$. Because $L$ is generally non-normal, left and right eigenvectors are distinct. We denote them by
\begin{equation}
\begin{aligned}
L\vect{x}_{\alpha}&=\lambda_{\alpha}\vect{x}_{\alpha},\\
\vect{y}_{\alpha}^{\dagger}L&=\lambda_{\alpha}\vect{y}_{\alpha}^{\dagger},
\end{aligned}
\label{eq:left-right-eig}
\end{equation}
and impose the biorthogonal normalization
\begin{equation}
\vect{y}_{\alpha}^{\dagger}\vect{x}_{\beta}=\delta_{\alpha\beta}.
\label{eq:biorthogonal}
\end{equation}
On the decaying sector one may then write the Jordan expansion
\begin{equation}
Q_0 e^{-Lt}Q_0
=
\sum_{\alpha:\,\abs{\lambda_{\alpha}}>\varepsilon_0}
e^{-\lambda_{\alpha} t}
\sum_{r=0}^{s_{\alpha}-1}
\frac{t^r}{r!}\,
\mathcal{N}_{\alpha,r},
\label{eq:jordan-expansion}
\end{equation}
where $s_{\alpha}$ is the size of the Jordan block associated with $\lambda_{\alpha}$ and $\mathcal{N}_{\alpha,r}$ is the corresponding nilpotent contribution. Here $\varepsilon_0$ is the numerical zero-mode threshold used later in the implementation.

\begin{figure*}[t]
    \centering
    \includegraphics[width=0.8\linewidth]{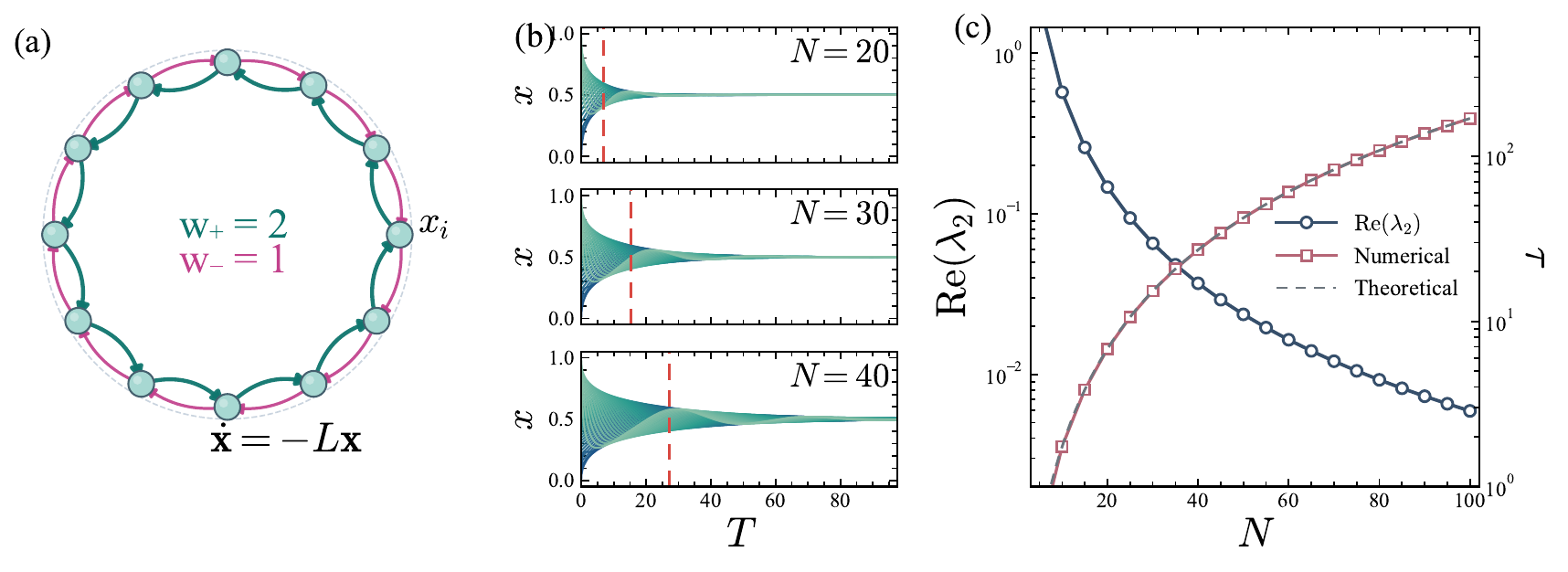}
    \caption{
    Spectral interpretation of the slow relaxation scale in a directed weighted ring. 
    (a) Minimal directed relaxation model on a ring with non-reciprocal weights, where the state evolves according to $\dot{x}=-Lx$. 
    (b) Time evolution of the node states for different system sizes, showing relaxation toward the stationary sector and progressively slower decay as $N$ increases. 
    (c) Size dependence of the slowest nonzero decay rate $\mathrm{Re}(\lambda_2)$ and the corresponding relaxation time $\tau$. 
    The measured relaxation time agrees with the spectral prediction $\tau \sim 1/\mathrm{Re}(\lambda_2)$, confirming that the real part of the slow branch controls the asymptotic decay scale.
    }
    \label{fig:overview}
\end{figure*}

Equation~\eqref{eq:jordan-expansion} implies that the asymptotic decay of the nonstationary dynamics is controlled by the smallest real part among the nonzero eigenvalues. More precisely, if
\begin{equation}
\mu
=
\min_{\abs{\lambda_{\alpha}}>\varepsilon_0}\ReL(\lambda_{\alpha}),
\label{eq:mu-def}
\end{equation}
then there exist constants $c_{\pm}>0$ and nonnegative integers $p_{\pm}$ such that
\begin{equation}
c_-\, t^{p_-} e^{-\mu t}
\le
\norm{Q_0 e^{-Lt}Q_0}
\le
c_+\, t^{p_+} e^{-\mu t}
\label{eq:metric-bounds}
\end{equation}
for sufficiently large $t$. The polynomial prefactors arise from possible Jordan blocks but do not affect the exponential rate. Substituting Eq.~\eqref{eq:metric-bounds} into Eq.~\eqref{eq:metric-definition} yields
\begin{equation}
\mathcal{M}(G)=\mu.
\label{eq:metric-equals-mu}
\end{equation}
Equation~\eqref{eq:mu-def} defines the controlling decay exponent, while Eq.~\eqref{eq:metric-equals-mu} identifies it with the robustness metric itself.

We therefore define $\lambda_2$ as the nonzero branch with smallest real part,
\begin{equation}
\lambda_2 \in \{\lambda_{\alpha}: \abs{\lambda_{\alpha}}>\varepsilon_0\},
\qquad
\ReL(\lambda_2)=
\min_{\abs{\lambda_{\alpha}}>\varepsilon_0}\ReL(\lambda_{\alpha}),
\label{eq:lambda2-def}
\end{equation}
and obtain the key identity
\begin{equation}
\mathcal{M}(G)=\ReL(\lambda_2).
\label{eq:metric-main}
\end{equation}
Equations~\eqref{eq:lambda2-def} and \eqref{eq:metric-main} provide the desired derivation: the robustness metric defined from long-time decay is exactly the real part of the slow nonzero branch.

When $\lambda_2$ is simple and spectrally isolated, the long-time dynamics reduces further to
\begin{equation}
Q_0 e^{-Lt}Q_0
=
e^{-\lambda_2 t}\,
\vect{x}_2\vect{y}_2^{\dagger}
\;+\;
\mathcal{O}\!\left(
e^{-(\ReL(\lambda_2)+\Delta)t}
\right),
\label{eq:dominant-branch}
\end{equation}
with $\Delta>0$ the real-part separation from the next branch. For each modal amplitude one has
\begin{equation}
a_2(t)\propto e^{-\lambda_2 t}
=
e^{-\ReL(\lambda_2)t}
e^{-i\operatorname{Im}(\lambda_2)t},
\label{eq:mode-decay}
\end{equation}
which provides the modal decomposition of the long-time dynamics. The factor $e^{-Re(\lambda_2)t}$ sets the asymptotic decay scale, while $e^{-i Im(\lambda_2)t}$ describes oscillatory modulation. For the relaxation metric defined in Eq.~\eqref{eq:metric-definition}, the relevant scalar quantity is therefore $ Re(\lambda_2)$.

This derivation clarifies a common modal-selection ambiguity in directed spectra. Ranking modes by modulus mixes dissipation with rotation, and treating the full complex eigenvalue as a scalar robustness index obscures the distinction between persistence and circulation. For the relaxation metric defined in Eq.~\eqref{eq:metric-definition}, the asymptotic scalar quantity is the smallest nonzero decay rate. For this long-time relaxation metric, the appropriate quantity is therefore $\mathrm{Re}(\lambda_2)$ rather than $|\lambda_2|$ or $\mathrm{Im}(\lambda_2)$.

Along an attack trajectory, let $m$ be the number of deleted nodes
and let $\lambda_\star^{(m)}$ denote the branch selected at that graph
state. The raw remaining ratio is defined as
\begin{equation}
\mathcal R(m)
=
\frac{
\max\left[
\mathrm{Re}\left(\lambda_\star^{(m)}\right),0
\right]
}{
\max\left[
\mathrm{Re}\left(\lambda_\star^{(0)}\right),0
\right]
},
\label{eq:remaining-ratio}
\end{equation}
which compares the surviving decay rate with its value before any deletion. It filters the entire zero eigenspace rather than assuming that the graph remains strongly connected, and it continues to make sense whenever a nonzero branch still exists away from the origin.

In the symmetric limit, the same construction reduces to the familiar undirected picture. If $A=A^{\mathsf{T}}$, then $L$ is real symmetric, all imaginary parts vanish, and the branch metric in Eq.~\eqref{eq:metric-main} becomes the Fiedler value. More explicitly,
\begin{equation}
\lambda_2
=
\min_{\vect{q}\perp \vect{1},\,\vect{q}\neq 0}
\frac{\vect{q}^{\mathsf{T}}L\vect{q}}
{\vect{q}^{\mathsf{T}}\vect{q}}
=
\frac{1}{2}
\min_{\vect{q}\perp \vect{1},\,\vect{q}\neq 0}
\frac{\sum_{i,j}w_{ij}(q_i-q_j)^2}
{\sum_i q_i^2},
\label{eq:fiedler-variational}
\end{equation}
which shows that the same metric becomes the weakest diffusive coupling channel in the reciprocal sector. The directed metric is therefore not an unrelated object; it can be viewed as an asymptotic-decay analogue of algebraic connectivity in the non-Hermitian setting.

The geometric meaning also becomes transparent in this limit. The Fiedler vector separates weakly communicating regions, and large edge differences across that vector mark the bottlenecks through which the slow mode is transmitted. Weighted-undirected networks fit the same logic with minimal change: symmetry keeps the branch real, while edge weights reshape which corridors dominate the variational minimum in Eq.~\eqref{eq:fiedler-variational}. Weighted-undirected networks should therefore be viewed as a spectrally filtered realization of the same principle rather than as a separate theory.

The spectral interpretation introduced above is illustrated in Fig.~\ref{fig:overview} using a directed relaxation process on a weighted ring. 
This example separates three ingredients of the robustness metric: the asymmetric operator, the decay of the state variables, and the scaling of the slow relaxation rate. 
The network in Fig.~\ref{fig:overview}(a) provides a minimal setting in which direction-dependent weights generate a non-reciprocal Laplacian while preserving a simple geometry. 
The time traces in Fig.~\ref{fig:overview}(b) show that perturbations decay toward the stationary sector, but the decay becomes increasingly slow as the system size grows. 
This slowing down is quantified in Fig.~\ref{fig:overview}(c), where the measured relaxation time follows the inverse scaling set by the slowest nonzero decay rate. 
Thus, even in a simple topology, the relevant robustness information is carried by the spectral relaxation scale rather than by connectivity alone.

The same example also motivates the use of $\mathrm{Re}(\lambda_2)$ as the branch metric. 
For a non-reciprocal Laplacian, the slow branch can in general be complex, but the long-time envelope of the projected dynamics is controlled by its real part. 
The numerical decay in Fig.~\ref{fig:overview}(b) and the agreement between the measured relaxation time and the spectral prediction in Fig.~\ref{fig:overview}(c) show that the real part of the slow branch provides the operative decay scale. 
This observation supplies the dynamical basis for the attack metric used below: node removal is evaluated by how strongly it reduces the branch-tracked relaxation rate.

\section{Modal susceptibility under node deletion}
\label{sec:algorithms}

This section derives the attack rule directly from the metric introduced in Sec.~\ref{sec:framework}. The relevant question is not which node looks most central, but which node produces the largest first-order reduction of the branch metric in Eq.~\eqref{eq:metric-main}. That question leads directly to MS. The remainder of the section turns the derivation into a complete adaptive protocol by specifying branch tracking, fallbacks, baselines, complexity, and the numerical ensembles.

Fix a graph state after $m$ deletions and let $\lambda_\star$, $\vect{x}$, and $\vect{y}$ denote the selected slow branch and its biorthogonal eigenvectors. For each candidate node $v$, introduce the coordinate-deletion matrix $P_v\in\mathbb{R}^{(N-1)\times N}$ and the coordinate mask
\begin{equation}
C_v=P_v^{\mathsf{T}}P_v=I-\vect{e}_v\vect{e}_v^{\mathsf{T}}.
\label{eq:coordinate-mask}
\end{equation}
Equation~\eqref{eq:coordinate-mask} isolates the survivor subspace associated with deleting node $v$.
The reduced Laplacian acting on the survivor space is
\begin{equation}
L^{(-v)}
=
P_v L P_v^{\mathsf{T}}
-
\diag\!\bigl(\{w_{iv}\}_{i\neq v}\bigr),
\label{eq:reduced-laplacian}
\end{equation}
because deleting node $v$ removes the coordinate itself and eliminates,
from every surviving row $i$, the term associated with the interaction
weight $w_{iv}$. Under the row-Laplacian convention, this reduces the
surviving diagonal entry $d_i$ by $w_{iv}$, while the off-diagonal
entries involving $v$ are removed by the coordinate projection. Lifting Eq.~\eqref{eq:reduced-laplacian} back to the original dimension gives the embedded survivor operator
\begin{equation}
\widetilde L^{(-v)}
=
C_vLC_v
-
\sum_{i\neq v}w_{iv}\,\vect{e}_i\vect{e}_i^{\mathsf{T}}.
\label{eq:embedded-reduced}
\end{equation}
Because node deletion is a finite, dimension-changing operation rather than an infinitesimal edge perturbation, the following quantity should be interpreted as a compressed first-order branch response rather than an exact derivative along a continuous perturbation path.

At the current attack step, let $n$ denote the number of surviving nodes, and let $\lambda_\star$, $\vect{x}$, and $\vect{y}$ denote the selected slow relaxation branch and its biorthogonal eigenvectors. An exact greedy node attack would test every candidate node, recompute the nonzero spectrum of $L^{(-v)}$, reselect the slow relaxation branch using the same branch-selection rule, and compare the new value of $\ReL(\lambda_\star)$. The modal-susceptibility strategy replaces that expensive operation by a compressed Rayleigh-type branch estimator,
\begin{equation}
\widehat{\lambda}_\star^{(-v)}
=
\frac{
(C_v\vect{y})^{\dagger}\widetilde L^{(-v)}(C_v\vect{x})
}{
(C_v\vect{y})^{\dagger}(C_v\vect{x})
}.
\label{eq:compressed-rayleigh}
\end{equation}
The predicted first-order branch drop is then defined as
\begin{equation}
\Delta_v^{(1)}
=
\lambda_\star-\widehat{\lambda}_\star^{(-v)}.
\label{eq:first-order-drop}
\end{equation}
Equations~\eqref{eq:compressed-rayleigh} and \eqref{eq:first-order-drop} form the starting point of the MS derivation.

The denominator in Eq.~\eqref{eq:compressed-rayleigh} follows directly from the biorthogonal normalization in Eq.~\eqref{eq:biorthogonal}:
\begin{equation}
(C_v\vect{y})^{\dagger}(C_v\vect{x})
=
\vect{y}^{\dagger}\vect{x}-y_v^{*}x_v
=
1-y_v^{*}x_v .
\label{eq:compressed-denominator}
\end{equation}
This factor is the normalization of the compressed slow branch. It becomes small when the deleted coordinate carries a large fraction of the biorthogonal modal weight.

We now turn to the numerator, where the effect of node deletion enters nontrivially. Using the embedded reduced operator in Eq.~\eqref{eq:embedded-reduced}, one obtains
\begin{equation}
(C_v\vect{y})^{\dagger}\widetilde L^{(-v)}(C_v\vect{x})
=
\vect{y}^{\dagger} C_v L C_v \vect{x}
-
\sum_{i\ne v} w_{iv} y_i^{*}x_i .
\label{eq:numerator-first-step}
\end{equation}
The first term is a compressed quadratic form. Expanding the projectors explicitly gives
\begin{align}
\vect{y}^{\dagger} C_v L C_v \vect{x}
&=
\vect{y}^{\dagger}
(I-e_v e_v^{\mathsf T})
L
(I-e_v e_v^{\mathsf T})
\vect{x}
\nonumber\\
&=
\vect{y}^{\dagger}L\vect{x}
-
\vect{y}^{\dagger}L e_v e_v^{\mathsf T}\vect{x}
\nonumber\\
&\quad
-
\vect{y}^{\dagger} e_v e_v^{\mathsf T}L\vect{x}
+
\vect{y}^{\dagger}e_v e_v^{\mathsf T}L e_v e_v^{\mathsf T}\vect{x}.
\label{eq:compressed-quadratic-expand}
\end{align}
Using the left--right eigenvalue relations in Eq.~\eqref{eq:left-right-eig}, Eq.~\eqref{eq:compressed-quadratic-expand} reduces to
\begin{equation}
\vect{y}^{\dagger} C_v L C_v \vect{x}
=
\lambda_\star
-
2\lambda_\star y_v^{*}x_v
+
L_{vv}y_v^{*}x_v .
\label{eq:compressed-quadratic-reduced}
\end{equation}
For the out-strength Laplacian convention used here, $L_{vv}=\sum_j w_{vj}$. Therefore, Eq.~\eqref{eq:numerator-first-step} becomes
\begin{align}
(C_v\vect{y})^{\dagger}\widetilde L^{(-v)}(C_v\vect{x})
&=
\lambda_\star
-
2\lambda_\star y_v^{*}x_v
\nonumber\\
&\quad
+
\left(\sum_j w_{vj}\right)y_v^{*}x_v
-
\sum_{i\ne v}w_{iv}y_i^{*}x_i .
\label{eq:compressed-numerator-algebraic}
\end{align}

To reveal the modal structure of this expression, we rewrite it in terms of edgewise differences. The $v$th component of the right eigenvalue equation gives
\begin{equation}
\sum_j w_{vj}(x_v-x_j)
=
\lambda_\star x_v ,
\label{eq:right-eig-v-component}
\end{equation}
while the $v$th component of the left eigenvalue equation gives
\begin{equation}
\sum_{i\ne v} w_{iv}y_i^{*}
=
\left(\sum_j w_{vj}-\lambda_\star\right)y_v^{*}.
\label{eq:left-eig-v-component}
\end{equation}
Substituting Eqs.~\eqref{eq:right-eig-v-component} and \eqref{eq:left-eig-v-component} into Eq.~\eqref{eq:compressed-numerator-algebraic}, the numerator can be recast as
\begin{align}
(C_v\vect{y})^{\dagger}\widetilde L^{(-v)}(C_v\vect{x})
&=
\lambda_\star(1-y_v^{*}x_v)
\nonumber\\
&\quad
-
\sum_j w_{vj} y_v^{*}(x_v-x_j)
\nonumber\\
&\quad
-
\sum_{i\ne v} w_{iv} y_i^{*}(x_i-x_v)
\nonumber\\
&\quad
+
\lambda_\star y_v^{*}x_v .
\label{eq:compressed-numerator}
\end{align}

Combining Eqs.~\eqref{eq:compressed-denominator} and \eqref{eq:compressed-numerator}, the predicted first-order drop of the tracked branch is
\begin{equation}
\Delta_v^{(1)}
=
\frac{
\sum_j w_{vj} y_v^{*}(x_v-x_j)
+
\sum_{i\ne v} w_{iv} y_i^{*}(x_i-x_v)
-
\lambda_\star y_v^{*}x_v
}{
1-y_v^{*}x_v
}.
\label{eq:drop-formula}
\end{equation}
The modal-susceptibility score is then defined by
\begin{equation}
Q_v^{\rm MS}
=
\ReL\!\left(\Delta_v^{(1)}\right).
\label{eq:qdir}
\end{equation}

Equation~\eqref{eq:drop-formula} has a direct physical interpretation. The first term quantifies the loss of outgoing modal support carried by node $v$, the second term accounts for the incoming support mediated by its neighbors, and the last term subtracts the direct modal contribution of the deleted coordinate itself. Together, these contributions estimate the reduction of the slow relaxation channel rather than a purely structural perturbation.

The denominator in Eq.~\eqref{eq:drop-formula} plays an equally important role. When $y_v^{*}x_v$ approaches unity, the deleted coordinate dominates the local normalization of the slow branch and the first-order estimate becomes ill conditioned. In practice, such cases are regularized by finite fallback scores to maintain a well-defined attack sequence.

When reciprocity is restored, the same construction reduces to the weighted Fiedler sector. Let $L=L^{\mathsf T}$ and let $\vect{u}$ be the normalized Fiedler vector, with $L\vect{u}=\lambda_2\vect{u}$, $\vect{u}^{\mathsf T}\vect{u}=1$, and $\vect{u}^{\mathsf T}\vect{1}=0$. After deleting node $v$, the compressed vector must be re-centered to remain orthogonal to the constant mode on the survivor graph:
\begin{equation}
\bar u_i
=
u_i+\frac{u_v}{n-1},
\qquad i\ne v .
\label{eq:compressed-fiedler-vector}
\end{equation}
Since $\sum_{i\ne v}u_i=-u_v$, this re-centered vector satisfies $\sum_{i\ne v}\bar u_i=0$, and its squared norm is
\begin{equation}
\sum_{i\ne v}\bar u_i^2
=
1-\frac{n}{n-1}u_v^2 .
\label{eq:compressed-fiedler-norm}
\end{equation}
The uniform re-centering in Eq.~\eqref{eq:compressed-fiedler-vector} does not change differences between surviving nodes. Thus, the removal of node $v$ changes the Fiedler Rayleigh quotient through the weighted incident edges and the normalization in Eq.~\eqref{eq:compressed-fiedler-norm}, yielding the symmetric specialization
\begin{equation}
Q_v^{\rm MS}
=
\frac{
\sum_{j\in\partial v} w_{vj}(u_v-u_j)^2
-
\lambda_2 \dfrac{n}{n-1}u_v^2
}{
1-\dfrac{n}{n-1}u_v^2
}.
\label{eq:qundirected}
\end{equation}
Equation~\eqref{eq:qundirected} shows that the undirected and weighted-undirected algorithms are symmetric specializations of the same branch-susceptibility principle.

\begin{figure}
    \centering
    \includegraphics[width=1\linewidth]{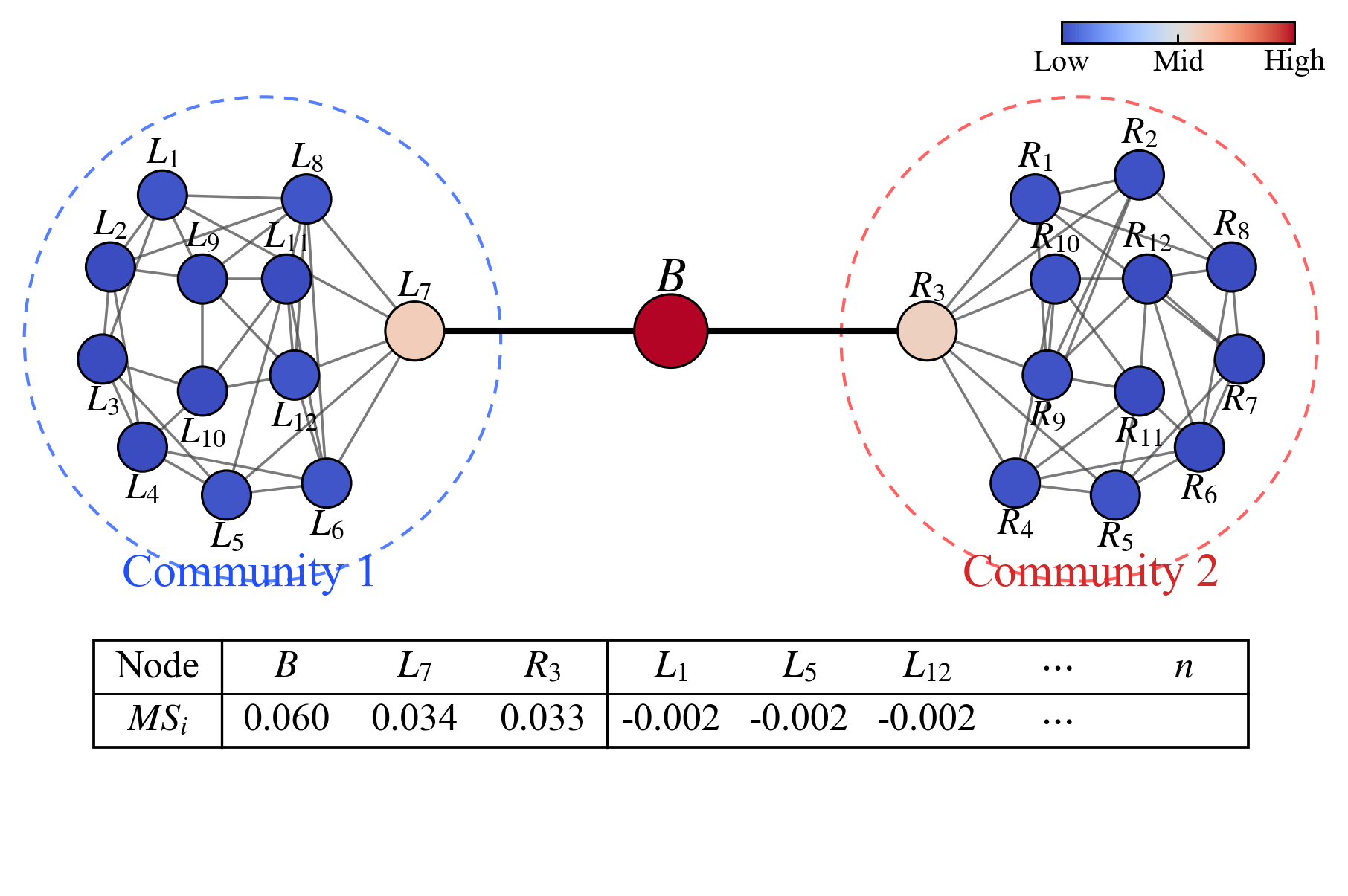}
    \caption{
    Illustration of the modal-susceptibility mechanism in a schematic two-community network. 
    The bridging node $B$ is important not merely because it is an inter-community bridge, but because it carries a large contribution to the branch-tracked slow-mode response. 
    Node colors represent the magnitude of modal susceptibility (MS), indicating the local contribution to the branch-tracked spectral response.
    }
    \label{fig:community}
\end{figure}

To make the mechanism underlying modal susceptibility more transparent, we consider a minimal schematic case in which a structural bridge also carries the dominant slow-mode load. This example is not meant to identify MS with structural bridge detection, but to illustrate how a structural motif becomes functionally vulnerable only when it supports the branch-tracked relaxation mode. As shown in Fig.~\ref{fig:community}, two densely connected communities are coupled through a single bridging node $B$. Although many nodes within each community may be structurally prominent, the slow collective mode is primarily transmitted through the inter-community channel mediated by $B$.

In this setting, removing node $B$ produces a disproportionately large reduction in the branch-tracked decay rate not simply because $B$ is a bridge, but because the slow mode is concentrated on the inter-community channel that it mediates. By contrast, removing nodes that are highly connected but confined within a single community has a much weaker effect on the same spectral branch. The relevant notion of importance is therefore not generic structural centrality, but the local support of the slow relaxation mode.

This observation clarifies the role of the modal-susceptibility construction. The score quantifies how strongly each node contributes to the branch-tracked response, as encoded by the biorthogonal eigenvectors of the slow mode. In particular, the branch-response estimator in Eq.~\eqref{eq:drop-formula} decomposes naturally into contributions associated with outgoing support, incoming support, and the direct modal weight of the node, thereby identifying the nodes that most effectively expose the slow branch.

The attack protocol also requires a rule for selecting the slow relaxation branch after each deletion. We first exclude numerical zero modes satisfying
\begin{equation}
|\lambda|\leq \epsilon_0,
\qquad
\epsilon_0=10^{-9}.
\label{eq:zero-threshold}
\end{equation}
At $m=0$, the selected branch is the nonzero eigenvalue with the smallest real part. At later steps, we first identify the nonzero eigenvalue closest to the previously selected branch,
\begin{equation}
\lambda_{\rm track}^{(m)}
=
\arg\min_{|\lambda_\alpha^{(m)}|>\epsilon_0}
\left|\lambda_\alpha^{(m)}-\lambda_\star^{(m-1)}\right|,
\label{eq:tracked-branch}
\end{equation}
and compare it with the current minimal-real-part branch $\lambda_{\rm min}^{(m)}$. The selected branch is
\begin{equation}
\lambda_\star^{(m)}
=
\begin{cases}
\lambda_{\rm track}^{(m)}, &
\ReL\!\left(\lambda_{\rm track}^{(m)}\right)
\leq
\ReL\!\left(\lambda_{\rm min}^{(m)}\right)+\delta_{\rm br},\\
\lambda_{\rm min}^{(m)}, & \text{otherwise},
\end{cases}
\label{eq:branch-selection}
\end{equation}
with $\delta_{\rm br}=10^{-6}$. This rule is not a purely analytic continuation of a single eigenvalue branch. Rather, it is a physically constrained tracking rule that preserves spectral continuity when possible while enforcing consistency with the slowest relevant decay rate.

Independent re-sorting can jump between distinct physical modes even when no dynamical crossover has occurred. Pure nearest-neighbor continuation can fail in the opposite direction by following a continuous branch that no longer controls the long-time decay. The combined rule in Eq.~\eqref{eq:branch-selection} balances spectral continuity against physical relevance and keeps the robustness curve tied to the decay metric derived in Sec.~\ref{sec:framework}.

The same logic explains why evaluation must not stop at the first loss of strong connectivity. As long as a nonzero branch survives away from the origin, the metric in Eq.~\eqref{eq:metric-main} remains meaningful. If biorthogonal normalization fails or no nonzero branch survives at a given step, the ranking temporarily falls back to total degree, or total strength in weighted graphs, but the deletion process continues and the branch history is reset rather than terminated. For cumulative attack evaluation, we use the monotone remaining-ratio
envelope
\begin{equation}
\widetilde R(m)
=
\min_{0\leq \ell\leq m}\mathcal R(\ell),
\label{eq:monotone-envelope}
\end{equation}
where $\mathcal R(m)$ is the raw branch-tracked remaining ratio.

This envelope measures the best degradation achieved up to step $m$, rather than asserting that the instantaneous slow-branch decay rate must be strictly monotone under node deletion. The raw branch trajectory $\mathcal{R}(m)$ can exhibit small upward shifts because node deletion changes both the operator dimension and the modal ordering. In the numerical comparisons below, the envelope in Eq.~\eqref{eq:monotone-envelope} is used only for the AUC-based degradation metric; the same spectral selection rule is applied to all methods.

We also evaluate an exact one-step diagnostic to validate the perturbative score. For graph state $G_m$ and candidate node $v$, define
\begin{equation}
\Delta^{\rm exact}_v(m)
=
\ReL\!\left(\lambda^{(m)}_\star\right)
-
\ReL\!\left(\lambda^{(m;-v)}_\star\right),
\label{eq:exact-drop}
\end{equation}
where $\lambda_\star^{(m;-v)}$ is recomputed after actually deleting node $v$. Comparing Eq.~\eqref{eq:exact-drop} with Eq.~\eqref{eq:qdir} tests whether the perturbative MS ranking approximates the exact greedy ordering.

There are three relevant complexity scales. The first is the present dense implementation of MS. 
If $n_m=N-m$ and $e_m$ are the surviving node and edge counts at step $m$, 
full spectral decomposition costs $\mathcal{O}(n_m^3)$ per step and node scoring adds $\mathcal{O}(e_m)$. This gives
\begin{equation}
T_{MS}^{\mathrm{dense}}(N)=\sum_{m=0}^{N-2}\mathcal{O}(n_m^3+e_m),
\label{eq:complexity-ms}
\end{equation}
which yields $\mathcal{O}(N^4)$ in the worst case. The memory cost is $\mathcal{O}(N^2)$ for storing dense matrices.

The second scale is a practical reduced complexity obtained by using branch-centered sparse partial eigensolvers with warm starts, such as shift-invert Arnoldi or inverse iteration near the tracked branch. If a partial solve costs $\mathcal{O}(k_{\mathrm{ev}} e_m)$ with $k_{\mathrm{ev}}$ Krylov or inverse-iteration steps, then one step of MS costs
\begin{equation}
\mathcal{O}(k_{\mathrm{ev}} e_m)+\mathcal{O}(e_m)
=
\mathcal{O}(k_{\mathrm{ev}} e_m),
\label{eq:complexity-ms-step-sparse}
\end{equation}
and the full attack scales as
\begin{equation}
T_{ MS}^{\mathrm{sparse}}(N)
=
\sum_{m=0}^{N-2}\mathcal{O}(k_{\mathrm{ev}} e_m)
=
\mathcal{O}(k_{\mathrm{ev}} N E),
\label{eq:complexity-ms-sparse}
\end{equation}
where $E$ is the initial edge count. Equation~\eqref{eq:complexity-ms-step-sparse} is the corresponding per-step cost, and Eq.~\eqref{eq:complexity-ms-sparse} is the full-trajectory cost. For sparse ensembles with bounded mean strength, Eq.~\eqref{eq:complexity-ms-sparse} reduces to $\mathcal{O}(k_{\mathrm{ev}}N^2)$.
The contrast between Eq.~\eqref{eq:complexity-ms} and Eq.~\eqref{eq:complexity-ms-sparse} separates the cost of the present dense implementation from the expected scaling of a sparse branch-centered implementation. 

The third scale is brute-force exact greedy node attack. At step $m$, one must evaluate all $n_m$ surviving candidate deletions and recompute the branch metric after each one. With dense diagonalization this gives
\begin{equation}
T_{\mathrm{brute}}^{\mathrm{dense}}(N)
=
\sum_{m=0}^{N-2}\mathcal{O}(n_m\cdot n_m^3)
=
\mathcal{O}(N^5),
\label{eq:complexity-brute-dense}
\end{equation}
whereas with branch-centered sparse partial solves one obtains
\begin{equation}
T_{\mathrm{brute}}^{\mathrm{sparse}}(N)
=
\sum_{m=0}^{N-2}\mathcal{O}(n_m k_{\mathrm{ev}} e_m).
\label{eq:complexity-brute-sparse}
\end{equation}
For sparse graphs with $e_m=\mathcal{O}(n_m)$, Eq.~\eqref{eq:complexity-brute-sparse} becomes $\mathcal{O}(k_{\mathrm{ev}}N^3)$. The dense expression in Eq.~\eqref{eq:complexity-brute-dense} and the sparse expression in Eq.~\eqref{eq:complexity-brute-sparse} both show that exact greedy node attack remains asymptotically more expensive than MS by one full factor of $N$.

For reference, the original FGIA edge attack achieves polynomial complexity $O(n^3+k(n^2+d_{\max}m))$ and dramatically improves on brute-force edge deletion \cite{Luo2026FGIA}. The node setting considered here is intrinsically harder because node deletion changes the dimension of the operator and removes all incident edges at once. This is precisely why Eq.~\eqref{eq:qdir}, rather than an edgewise aggregation alone, is needed.

The numerical study uses three canonical ensembles: scale-free, small-world, and random graphs. Directed runs use the non-Hermitian formulation in Eqs.~\eqref{eq:drop-formula} and \eqref{eq:qdir}, whereas the undirected and weighted-undirected runs use the symmetric specialization in Eq.~\eqref{eq:qundirected}. These three ensembles stress distinct mechanisms of vulnerability: heterogeneity and localization in scale-free graphs, shortcut-mediated transport in small-world graphs, and distributed modal mixing in random graphs. A method that remains effective across all three cases is therefore plausibly responding to the slow branch itself rather than to one special mesoscopic motif.

\section{Results}
\label{sec:results}
This section first presents the directed results and then examines the
weighted-undirected symmetric limit. We begin with three directed
ensembles that motivate the non-Hermitian formulation, then show how
the same branch-susceptibility principle reduces to the symmetric
Fiedler sector, and finally test finite-size trends and one-step
prediction accuracy. Throughout, the emphasis is not on isolated
numerical margins, but on a consistent physical pattern: MS most
strongly degrades the tracked slow branch, or remains statistically
comparable to the strongest spectral baseline, because it targets the
spectral carrier of long-time relaxation rather than a generic
structural proxy.

For every attack figure in this paper, the horizontal axis is the
removed-node fraction
\begin{equation}
f=\frac{m}{N-1},
\label{eq:axis-x}
\end{equation}
where $m$ is the number of deleted nodes. The vertical axis is the
monotone remaining branch-robustness ratio
\begin{equation}
\widetilde{R}(f)\equiv \widetilde{R}(m),
\label{eq:axis-y}
\end{equation}
where $\widetilde{R}(m)$ is the envelope of the raw remaining ratio
defined by Eqs.~\eqref{eq:remaining-ratio} and
\eqref{eq:monotone-envelope}. Smaller curves therefore correspond to
stronger attacks, because they indicate a faster collapse of the
branch-tracked relaxation metric. Equations~\eqref{eq:axis-x} and
\eqref{eq:axis-y} define the axes used in all comparison figures.

The scalar summary used in the lower panels is the area under the
remaining-ratio curve,
\begin{equation}
\mathrm{AUC}
=
\int_0^1 \widetilde{R}(f)\,\mathrm{d} f
\approx
\frac{1}{N-1}
\sum_{m=0}^{N-2}
\frac{\widetilde{R}(m+1)+\widetilde{R}(m)}{2}.
\label{eq:auc-def}
\end{equation}
Because the ordinate measures remaining robustness rather than damage,
smaller AUC indicates a more effective attack. Equation~\eqref{eq:auc-def}
is the cumulative degradation measure compared across methods throughout
this section.

\begin{figure*}[t] 
    \centering
    \includegraphics[width=1\linewidth]{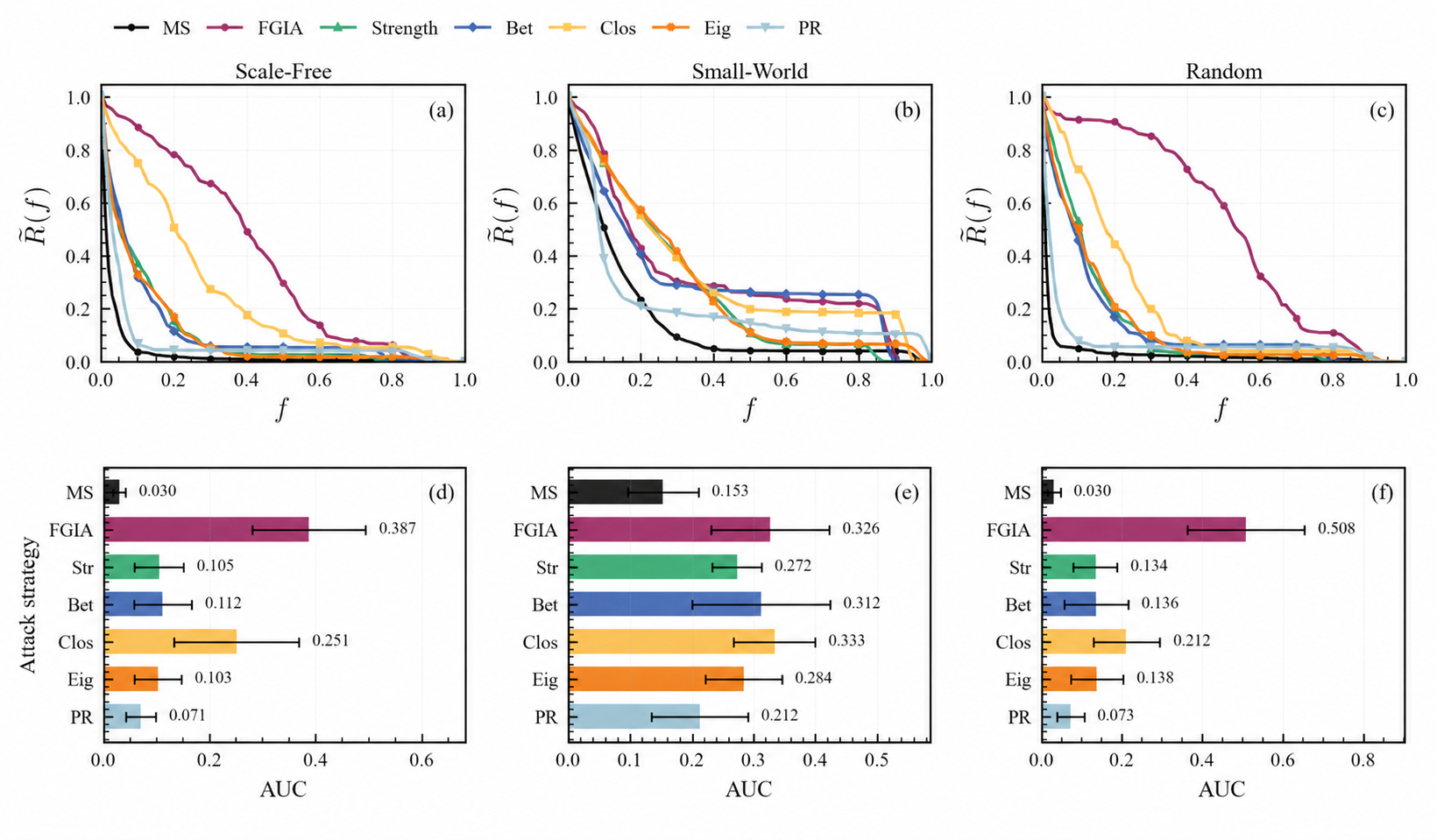}
    
\caption{Branch-aware attack performance in directed networks.
Top row: monotone remaining branch-robustness ratio $\widetilde R(f)$
under adaptive node removal.
Bottom row: AUC summary of cumulative degradation.
Curves and bars are averaged over 20 realizations, with error bars denoting standard deviations.
All networks have $N=80$ nodes.
Scale-free, small-world, and random ensembles are generated using BA($m=3$), WS($k=6,p=0.15$), and directed ER($p=0.075$), respectively.
BA and WS edges are replaced by oppositely oriented weighted edges; ER graphs are generated directly as directed graphs.
All nonzero weights are drawn from $U[0.5,2]$.
}
    \label{fig:directed}
\end{figure*}

\subsection{Directed ensembles: branch-sensitive vulnerability}
\label{subsec:directed-results}

We first compare MS with both structural baselines and a representative spectral-perturbation baseline. 
The structural baselines include total strength, betweenness, closeness, eigenvector centrality, and PageRank, all recomputed adaptively after each deletion. 

As a representative spectral-perturbation baseline, we use an
FGIA-type induced-node ranking. FGIA and related Fiedler-contribution
methods represent a class of ranking algorithms derived from
first-order spectral perturbation theory, in which the importance of
an edge is estimated from its contribution to a selected slow
Laplacian mode or to algebraic connectivity
\cite{Luo2026FGIA,Jiang2024FiedlerContribution,Zhou2024ConnectivityRankIndex,Hu2025AcceleratingSynchronization,Jiang2023KeyCycles,wen2025dynamics}.
These methods are naturally formulated for edge modification, since
deleting or weakening an edge preserves the dimension of the
Laplacian. Because the present problem concerns node removal rather
than edge removal, we convert the edge-level spectral score into a
node-level score by aggregating the incident FGIA-type contributions.

Let $g_{ij}^{\rm FGIA}$ denote the current edge score assigned by the
FGIA-type spectral perturbation rule. The corresponding induced-node
score is defined as
\begin{equation*}
Q_v^{\rm FGIA}
=
\sum_j g_{vj}^{\rm FGIA}
+
\sum_i g_{iv}^{\rm FGIA}.
\end{equation*}
This construction preserves the edgewise spectral information provided
by first-order perturbation theory while producing a complete adaptive
node ranking. It should therefore be interpreted as a representative
spectral-perturbation ranking baseline for node attack, rather than as
a response theory specifically derived for dimension-changing vertex
deletion.

For all methods, the attack is adaptive: after each deletion, the network, the spectral observable, and the corresponding node scores are recomputed. 
In the directed case, closeness is evaluated on the reversed digraph to preserve the outward-reach interpretation used in the implementation, and eigenvector-centrality failures fall back to strength for that step only.  Every method produces a complete deletion order of length $N-1$, and all curves are evaluated using the same branch-tracked spectral observable in Eq.~\eqref{eq:remaining-ratio}. 
Thus, the comparison tests the spectral information content of each ranking rather than mixing adaptive and nonadaptive protocols.

The directed calculations show a consistent qualitative pattern: among the adaptive strategies considered here, MS produces the strongest degradation of the tracked spectral branch across all three ensembles in Fig.~\ref{fig:directed}. 
While the margin varies with ensemble class, MS remains the best-performing method or clearly separated from nonspectral structural baselines. 
This robustness across structurally distinct systems indicates that the advantage of MS is not tied to a specific topology, but to its direct targeting of the slow spectral branch.

The two-row structure of Fig.~\ref{fig:directed} clarifies this point. 
The upper panels quantify how rapidly the slow mode is weakened during adaptive deletion, whereas the lower panels summarize the cumulative degradation through the AUC. 
Read together, they separate instantaneous sensitivity from integrated robustness loss. 
In both views, the comparison is performed on the same spectral observable, ensuring that the evaluation reflects persistence of the slow branch rather than generic structural damage.

The mechanism underlying this behavior is common across ensembles. 
MS ranks nodes according to their biorthogonal support on the selected slow branch, thereby identifying the modal-load carriers of the dominant decay channel. 
Removing such nodes directly disrupts the pathways through which the slow relaxation branch propagates, rather than indirectly affecting connectivity, reachability, or static centrality.

The specific manifestation of this mechanism depends on network organization. 
In scale-free networks, where strength heterogeneity creates a dense core and sparse periphery, the slow branch tends to localize on nodes that mediate communication between these regions. 
MS therefore selects hub-bridges that couple the core to branch-sensitive peripheral sectors, rather than hubs confined within the core. 
In small-world networks, the slow branch is shaped by the interplay between local order and long-range shortcuts. 
MS preferentially targets shortcut junctions that control inter-regional transport, even when these nodes are not structurally dominant. 
In random networks, where clear modular structure is absent, the slow branch is supported by a distributed set of nodes spanning multiple weakly organized regions. 
In this case, MS identifies modal mixing centers that mediate global amplitude exchange. 
These differences reflect distinct realizations of the same spectral principle rather than separate mechanisms.

This mechanism-level interpretation also explains the systematic gap between MS and the baseline methods. 
Strength and betweenness emphasize structural prominence, while closeness and PageRank capture reachability or recursive importance. 
These measures can identify structurally influential nodes, but they do not directly measure how strongly a node supports the selected slow relaxation branch. 
Consequently, they may remove nodes that are important for graph organization without producing the largest reduction in the branch-tracked decay rate.

The comparison with FGIA is more subtle because FGIA is already spectral. Its weaker performance in the directed node-attack setting does not mean that spectral perturbation is uninformative; rather, it reflects a mismatch between the perturbation object and the damage operation. 

FGIA is derived from an edge-level perturbation perspective, where deleting or modifying an edge preserves the dimension of the Laplacian and naturally fits a fixed-dimensional spectral-sensitivity calculation. Node removal is different: it deletes a coordinate, removes all incident edges, and changes the diagonal normalization of the survivor operator. This dimension-changing compression cannot be fully captured by summing edgewise contributions. MS directly approximates the first-order response of the branch-tracked robustness under this vertex-level compression, which aligns the ranking objective with the evaluation quantity.

Another important observation is that the advantage of MS persists beyond the first loss of strong connectivity. In directed systems, a nonzero decay branch can remain dynamically meaningful even after topological breakup. Continuing the evaluation in this regime reveals a clear separation between structural disintegration and spectral weakening. The sustained performance of MS therefore confirms that it captures a genuinely dynamical notion of vulnerability tied to the slow relaxation mode.

Overall, the directed results support a unified physical picture: vulnerability is governed by the nodes that carry the slow spectral branch. 
By directly targeting this branch, MS provides a mechanism-driven ranking that remains effective across heterogeneous directed networks.

\begin{figure*}[t] 
    \centering
    \includegraphics[width=1\linewidth]{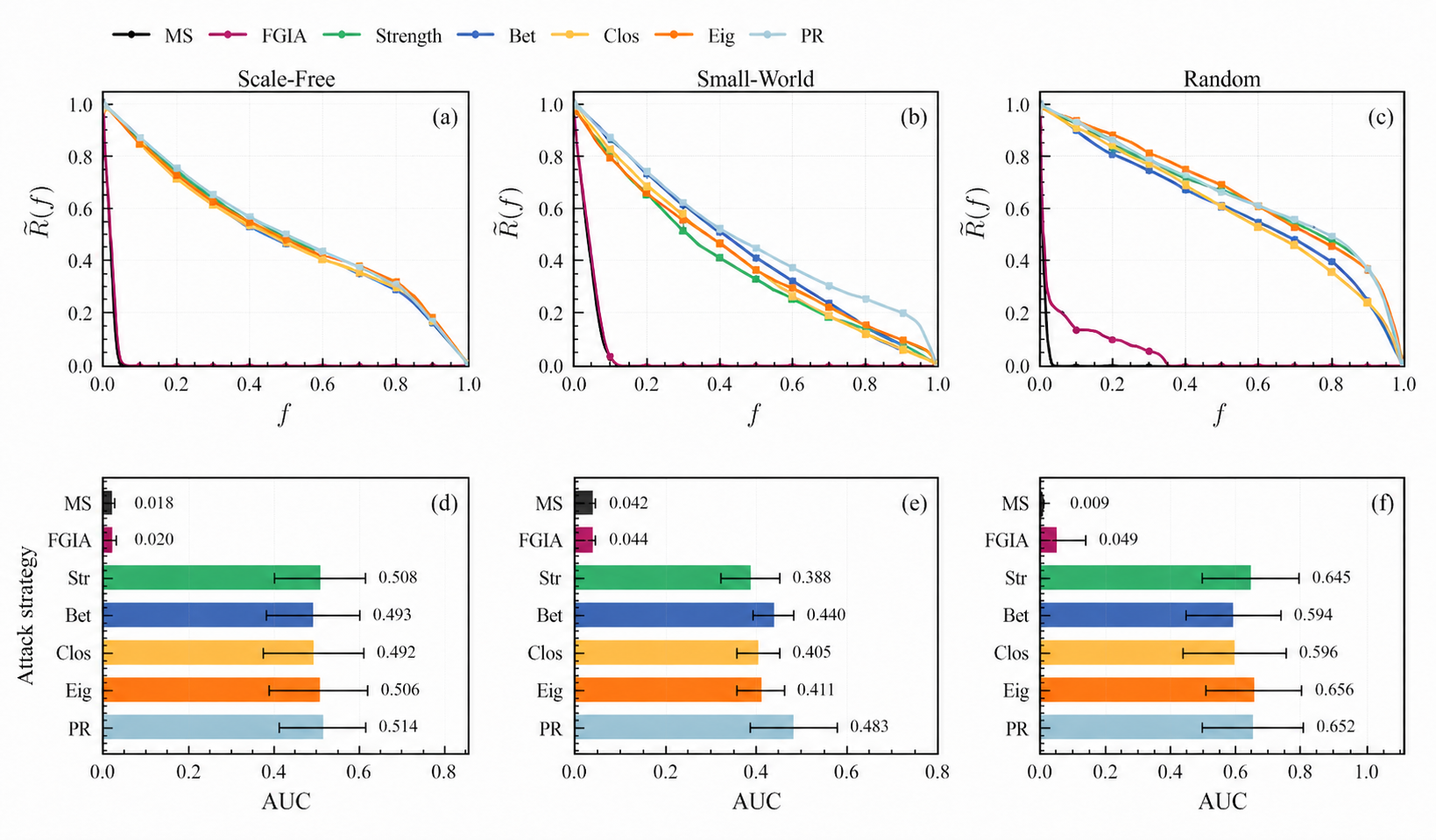}
    
\caption{
Branch-aware attack performance in the symmetric weighted-undirected limit.
Top row: monotone remaining branch-robustness ratio $\widetilde R(f)$
under adaptive node removal.
Bottom row: AUC summary of cumulative degradation.
Curves and bars are averaged over 20 realizations, with error bars denoting standard deviations.
All networks have $N=80$ nodes and are generated using BA($m=3$), WS($k=6,p=0.15$), and ER($p=0.08$).
Edges are undirected, with weights drawn from $U[0.5,2]$.
In this reciprocal limit, the tracked slow branch $\lambda_\star$ reduces to the weighted Fiedler eigenvalue $\lambda_2$.
}
    \label{fig:undirected}
\end{figure*}

\subsection{Symmetric limit: weighted-undirected networks}
\label{subsec:symmetric-results}

The weighted-undirected experiments test whether the modal-susceptibility principle remains valid once reciprocity is restored while heterogeneous edge weights are retained. 
In this symmetric limit, the slow spectral branch becomes purely real and reduces to the weighted Fiedler sector, providing a clearer geometric interpretation of the same robustness mechanism. 

The main qualitative result is largely preserved: MS produces the fastest degradation of the branch-tracked robustness or remains statistically close to the strongest spectral baseline across the weighted-undirected ensembles. FGIA becomes a particularly close competitor in this limit because both methods probe the same real Fiedler sector once non-Hermitian effects are removed.

This persistence is important because it shows that MS is not merely exploiting non-Hermitian spectral effects; it also captures the node-level compression of the slow mode in the reciprocal limit. Rather, it reflects a more general principle: node importance is determined by how strongly a node supports and perturbs the slow mode. The symmetric results therefore provide a controlled limit in which the same mechanism can be interpreted in terms of Fiedler-mode geometry.

In this limit, the role of MS becomes more transparent. In scale-free networks, it selects interfacial nodes that mediate communication between dense and sparse regions. In small-world networks, it targets shortcut bottlenecks that connect otherwise weakly coupled sectors. In random networks, it favors distributed relays spanning multiple regions. When weights are introduced, the same structure persists but is filtered by spectral importance: nodes are selected not by the number of incident links alone, but by their contribution to the weighted channels carrying the slow mode. These patterns confirm that MS tracks modal load rather than structural prominence.

The comparison with FGIA is particularly instructive. In symmetric spectra, FGIA provides a strong spectral baseline by ranking edges according to their contribution to the Fiedler mode. When these edge sensitivities are aggregated to nodes, the resulting score can approximate the effect of node removal and, in some realizations, can approach the performance of MS. This behavior is expected because both methods probe the same real slow branch once non-Hermitian effects are absent.

However, this agreement is only partial. FGIA is derived from a fixed-dimensional edge-perturbation framework: deleting an edge preserves the size of the Laplacian and therefore fits naturally into a standard spectral-sensitivity calculation. Node removal is different. It deletes a coordinate together with all incident edges and induces a dimension-changing compression of the operator. This compression cannot be fully captured by simply summing edge scores. Thus, FGIA remains an indirect node-level proxy, whereas MS directly approximates the first-order response of the branch-tracked robustness under vertex deletion.

This distinction explains why MS can retain an advantage in the symmetric setting and why its advantage becomes more pronounced in directed networks. The gain of MS does not come from exploiting a special topology, but from resolving the correct spectral response mechanism associated with node removal: node-level vulnerability is governed by operator compression rather than edgewise perturbation.

Taken together with the directed results, the symmetric experiments complete the unification picture. The directed case represents the generic non-Hermitian setting, while the undirected and weighted-undirected cases are symmetry-restored limits of the same response theory. Across all cases, vulnerability is governed by the nodes that carry the slow spectral branch, with the apparent structural role---bridge, shortcut, or relay---emerging from how the mode is supported in each network.

To further verify that this ordering is not limited to the benchmark ensembles, 
Table~\ref{tab:filtered-runnable-ms-role-map} extends the comparison to a broader set of real-world networks. 
Despite substantial differences in size, topology, directionality, and weighting, the same qualitative pattern persists: 
MS achieves the lowest or near-lowest AUC values in most datasets. 
This cross-dataset consistency provides compact evidence that branch-aware susceptibility captures a general and robust notion of network vulnerability. 
All empirical datasets used in Table~\ref{tab:filtered-runnable-ms-role-map} are publicly available from the Network Repository~\cite{nr}.

\begin{table*}
\centering
\caption{
AUC comparison across real-world networks.
Smaller values indicate more effective degradation of the branch-tracked robustness.
For each network, the best-performing method (minimum AUC) is shown in bold. 
}
\setlength{\tabcolsep}{7pt}
\begin{tabular}{lcccc|ccccccc}
\hline
Name & $n$ & $m$ & W & D & MS & FGIA & Str& Bet & Clos & Eig & PR \\
\hline

\multicolumn{12}{l}{\textit{Animal Networks}} \\
moreno\_mac & 62 & 1187 & Y & Y & \textbf{0.0082} & 0.1230 & 0.4180 & 0.4836 & 0.8443 & 0.0246 & \textbf{0.0082} \\
moreno\_cattle & 28 & 217 & Y & Y & \textbf{0.0408} & 0.8333 & 0.6111 & 0.3519 & 0.4821 & 0.3519 & 0.9444 \\
moreno\_rhesus & 16 & 111 & Y & Y & \textbf{0.1175} & 0.7387 & 0.3630 & 0.6649 & 0.3448 & 0.4172 & 0.3977 \\

\multicolumn{12}{l}{\textit{Dynamic Networks}} \\
copresence-InVS13 & 95 & 3915 & N & N & \textbf{0.0576} & 0.9947 & 0.3712 & 0.4605 & 0.4772 & 0.3581 & 0.3724 \\

\multicolumn{12}{l}{\textit{Ecology Networks}} \\
eco-foodweb-baydry & 128 & 2106 & N & Y & \textbf{0.0155} & 0.9414 & 0.6407 & 0.5797 & 0.6797 & 0.7006 & 0.6204 \\
eco-mangwet & 97 & 1446 & N & N & \textbf{0.0207} & 0.8711 & 0.5442 & 0.5380 & 0.4270 & 0.5443 & 0.5113 \\
eco-stmarks & 54 & 350 & N & N & \textbf{0.0302} & 0.8611 & 0.2087 & 0.1944 & 0.2326 & 0.2139 & 0.2540 \\
eco-florida & 128 & 2075 & N & N & \textbf{0.0415} & 0.9414 & 0.6120 & 0.5172 & 0.5172 & 0.7207 & 0.6280 \\

\multicolumn{12}{l}{\textit{Interaction Networks}} \\
email-enron-only & 143 & 623 & N & N & \textbf{0.0035} & \textbf{0.0035} & 0.8350 & 0.8111 & 0.8455 & 0.9289 & 0.8502 \\
convote & 219 & 586 & Y & Y & \textbf{0.0393} & 0.0986 & 0.1030 & 0.0866 & 0.1358 & 0.3142 & 0.2200 \\

\multicolumn{12}{l}{\textit{Social Networks}} \\
moreno\_highschool & 70 & 366 & Y & Y & \textbf{0.0105} & 0.0976 & 0.1093 & 0.0730 & 0.2454 & 0.1490 & 0.0691 \\
moreno\_innovation & 241 & 1098 & N & Y & \textbf{0.0237} & 0.3060 & 0.3394 & 0.2526 & 0.5362 & 0.3861 & 0.4760 \\
moreno\_seventh & 29 & 376 & Y & Y & \textbf{0.0616} & 0.9464 & 0.2857 & 0.1398 & 0.1843 & 0.3856 & 0.3338 \\
librec-filmtrust-trust & 874 & 1853 & N & Y & \textbf{0.0534} & 0.1929 & 0.0711 & 0.0969 & 0.1112 & 0.0721 & 0.1475 \\

\hline
\end{tabular}
\label{tab:filtered-runnable-ms-role-map}
\end{table*}
\subsection{Finite-size scaling and validation}
\label{subsec:robustness}

For the real-world networks in Table~\ref{tab:filtered-runnable-ms-role-map}, self-loops are removed and multiple edges, when present, are aggregated into a single weighted edge. 
Unweighted networks are assigned unit edge weights, whereas weighted networks use their reported weights after restricting to nonnegative entries. 
The attack and evaluation are both performed on the largest weakly connected component. 
For directed networks, the non-Hermitian slow-branch rule described above is used; for undirected networks, the symmetric Fiedler specialization is used. 
Zero modes are filtered using the same threshold $\epsilon_0 = 10^{-9}$, and all methods are evaluated using the same branch-tracked remaining ratio and AUC.

Notably, in some undirected networks, FGIA fails to fully degrade the slow branch, whereas MS maintains its optimal performance, highlighting its robustness advantage. 
This difference reflects the intrinsic limitation of edge-based spectral perturbation in capturing vertex-level dimension-changing effects: 
while FGIA aggregates edge sensitivities, MS directly approximates the first-order response of the branch-tracked slow mode under node removal, 
ensuring consistent ranking even when network heterogeneity or local modal localization is strong.

The preceding ensemble comparisons show that MS efficiently degrades the branch-tracked relaxation metric across directed and weighted-undirected networks. We now examine whether this performance reflects a robust branch-sensitivity mechanism rather than a finite-size or single-parameter effect. Three complementary diagnostics are used. First, we measure the finite-size dependence of the AUC gap between MS and the strongest competing adaptive baseline. Second, we compare the perturbative MS prediction with the exact one-step branch drop obtained by deleting each candidate node and recomputing the selected slow branch. Third, we test the stability of this approximation as the average out-strength is varied. These checks directly assess whether MS acts as an efficient surrogate for exact greedy node removal.

\begin{figure}[t] 
    \centering
    \includegraphics[width=0.8\linewidth]{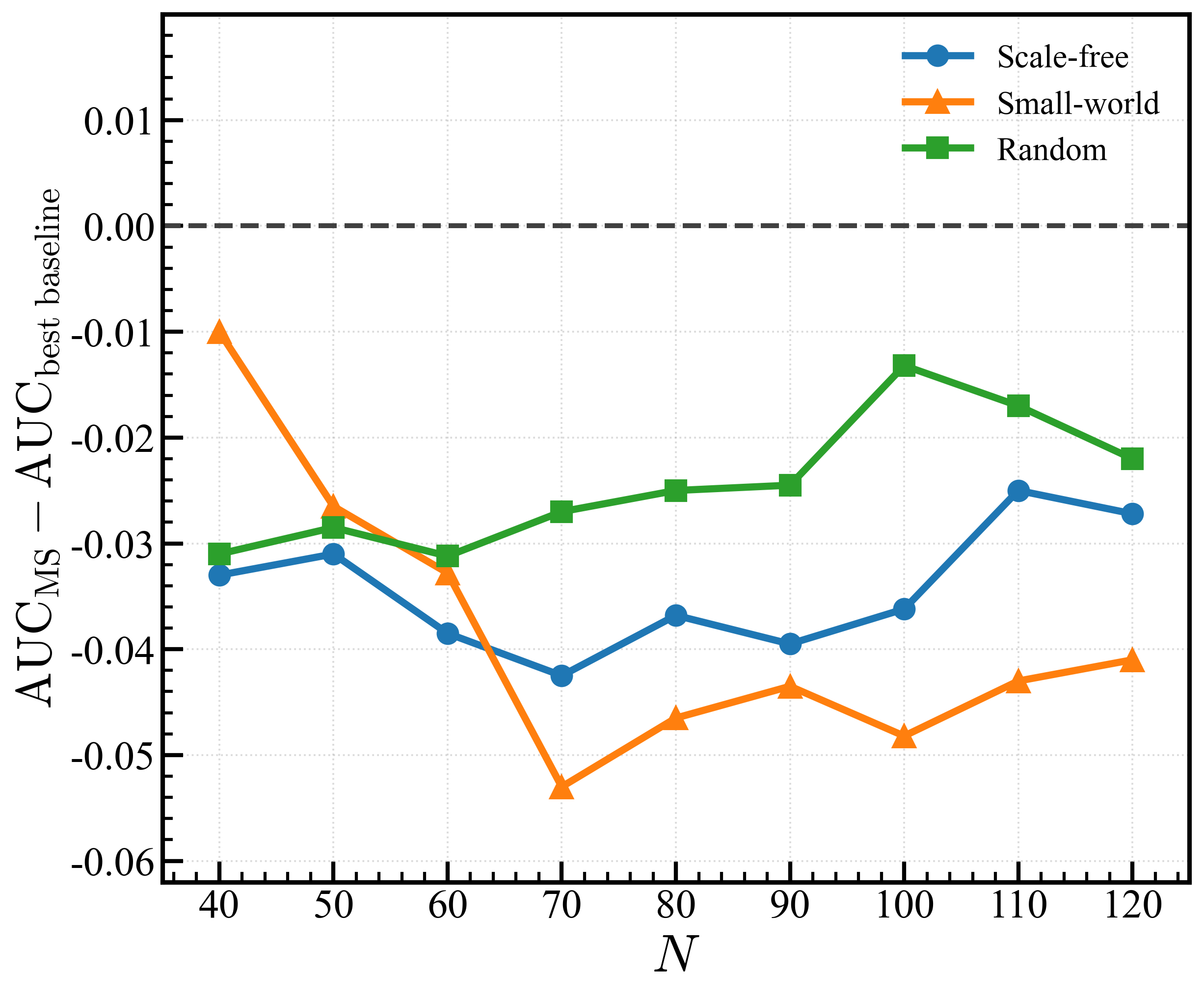}
    \caption{
Finite-size scaling of the AUC gap between MS and the strongest competing adaptive baseline across scale-free, small-world, and random ensembles.
Each data point is averaged over 20 independent realizations.
The gap is defined as
\(\Delta_{\rm AUC}={\rm AUC}_{\rm MS}-\min_{\mathcal{B}\neq{\rm MS}}{\rm AUC}_{\mathcal{B}}\).
Negative values indicate that MS achieves lower AUC than the strongest competing baseline, corresponding to faster degradation of the branch-tracked robustness.
}
    \label{fig:aucgap}
\end{figure}

The finite-size trends are summarized in Fig.~\ref{fig:aucgap}. The quantity
\(\Delta_{\rm AUC}\) compares MS with the best non-MS baseline at the same system size and in the same ensemble. Negative values therefore indicate that MS produces a smaller remaining-robustness AUC than all competing adaptive strategies. Across the tested sizes, the gap remains predominantly negative or close to zero, showing that the advantage of MS is not restricted to a single network size. The magnitude of the gap varies with ensemble class, as expected: heterogeneous scale-free networks and shortcut-mediated small-world networks provide more structured slow modes, whereas random networks yield more distributed modal support and consequently smaller separations between methods. This behavior supports the interpretation that MS responds to the organization of the slow branch rather than to a size-specific structural artifact.

A more direct test of the perturbative construction is obtained by comparing the MS-predicted branch drop with the exact one-step removal. For a graph state \(G_m\) and a candidate node \(v\), the exact drop is computed by deleting \(v\), recomputing the selected nonzero branch, and evaluating
\[
\Delta_v^{\rm exact}
=
\ReL\!\left(\lambda_\star^{(m)}\right)
-
\ReL\!\left(\lambda_\star^{(m;-v)}\right).
\]
This quantity represents the ideal greedy information available at one step, but obtaining it for all candidates requires repeated spectral recomputation. The MS score \(Q_v^{\rm MS}\), by contrast, estimates the same response from the current slow branch and its biorthogonal eigenvectors.

\begin{figure}[t]
    \centering
    \includegraphics[width=1\linewidth]{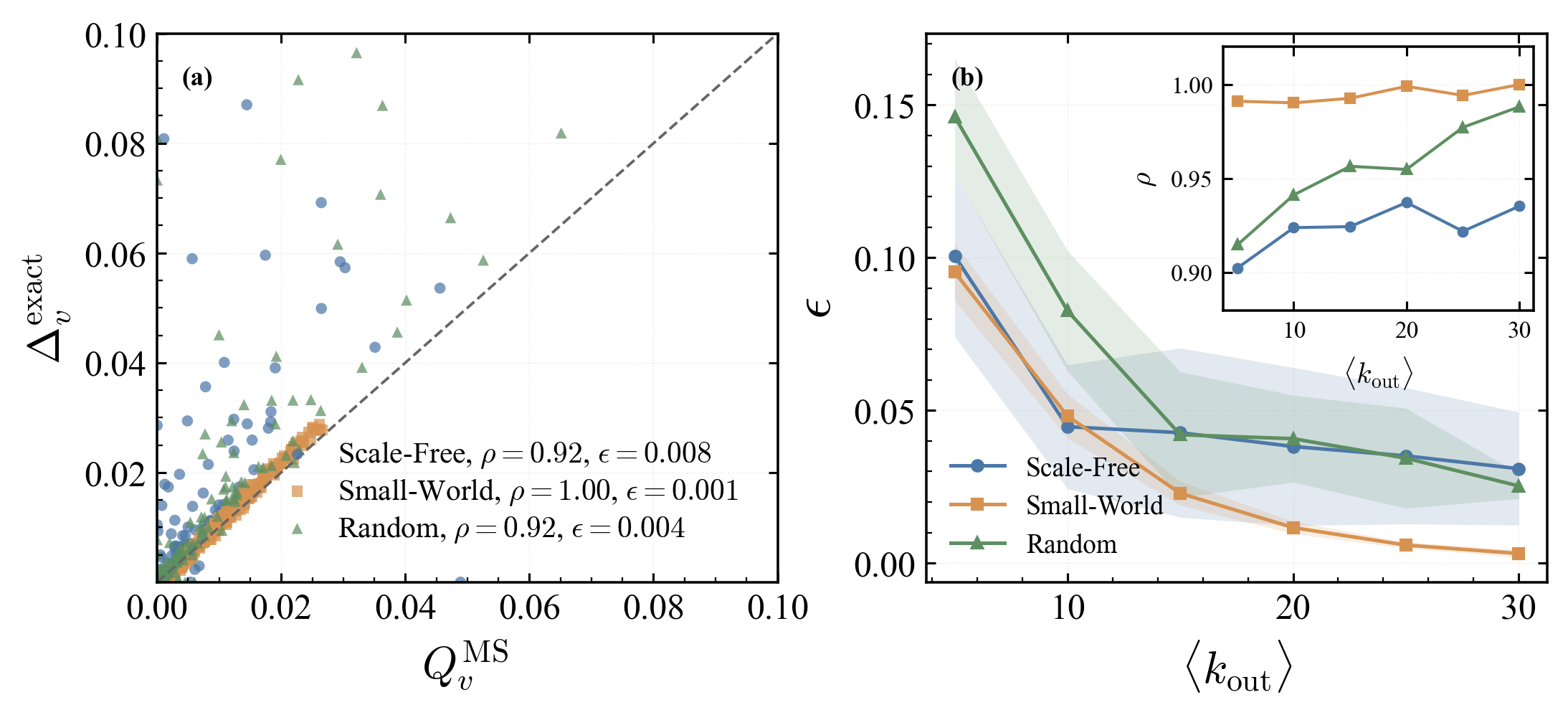}
\caption{
Validation of modal susceptibility (MS) in directed networks.
(a) Comparison between the MS-predicted branch drop \(Q_v^{\rm MS}\) and the exact one-step drop \(\Delta_v^{\rm exact}\), obtained by deleting each candidate node and recomputing the selected slow branch.
The diagonal indicates perfect agreement.
(b) Normalized prediction error \(\epsilon\) as a function of the average out-strength, with each point averaged over 100 independent realizations.
Inset: Spearman rank correlation \(\rho\), also averaged over 100 realizations.
}
    \label{fig:CompareExact}
\end{figure}

The comparison in Fig.~\ref{fig:CompareExact}(a) shows that the perturbative prediction captures the exact one-step response with high accuracy in small-world networks, where the points lie close to the diagonal. This agreement is consistent with the spectral structure of these graphs: the slow branch is typically supported over a broad set of nodes, so deleting one node does not strongly violate the first-order approximation. The branch separation is also more regular, which further stabilizes the compressed response estimate.

Larger deviations appear in scale-free and random networks. In scale-free graphs, the slow branch can be more localized because of the coexistence of dense hub regions and sparse peripheral sectors. Deleting a node with strong localized support can therefore induce higher-order changes that are not fully captured by the first-order MS formula. Random networks show weaker but still visible deviations, reflecting fluctuations in local connectivity and in the spectral separation of the slow branch. These deviations affect the predicted magnitude of the drop, but they do not imply a breakdown of the ranking mechanism.

The ranking accuracy is quantified by the Spearman correlation shown in the inset of Fig.~\ref{fig:CompareExact}(b). The correlation remains close to unity across the tested ensembles and densities, indicating that MS preserves the node ordering associated with the exact greedy drop even when the absolute response is not perfectly reproduced. This distinction is important for adaptive attack: the relevant task is to select the most damaging node at each step, not necessarily to predict the exact value of the branch reduction. From this perspective, MS provides a computationally efficient approximation to exact greedy node removal while avoiding the additional factor of candidate-wise spectral recomputation.

The density dependence in Fig.~\ref{fig:CompareExact}(b) further supports this interpretation. The normalized prediction error \(\epsilon\) remains bounded over the tested range of average out-strength and generally decreases as the network becomes denser. Denser graphs tend to produce less singular node-removal responses because the slow branch is distributed over more alternative pathways. Small-world networks maintain the lowest errors, consistent with their relatively stable slow-mode structure. Scale-free and random networks exhibit larger fluctuations, showing that the accuracy of MS is controlled not only by density but also by modal localization and spectral separation.

Together, the finite-size and exact-drop diagnostics support the central approximation behind MS. The method is not an exact greedy attack, nor is it intended to be one. Rather, it is a branch-aware first-order surrogate that uses the current slow mode to estimate which node most strongly reduces the long-time decay rate. Its effectiveness is therefore expected to be highest when the tracked branch is spectrally separated and moderately delocalized, and less exact when node removal triggers strong localization effects or branch reorganization. The numerical tests show that, despite these limitations, MS retains high ranking fidelity and consistently captures the nodes most relevant to the branch-tracked relaxation response.

\section{Discussion}

The central result of this work is that relaxation robustness in
networked dynamical systems is controlled by a spectral object rather
than by connectivity alone. For the projected decaying dynamics, the
relevant quantity is the real part of the selected slow nonzero branch,
$\mathrm{Re}(\lambda_\star)$, which sets the asymptotic decay scale of
the nonstationary sector. Node importance should therefore be assessed
by how strongly a node supports or perturbs this branch, not only by
its topological prominence.

This distinction is especially important in directed networks. Once
reciprocity is broken, the Laplacian becomes non-Hermitian, the slow
branch can be complex, and left and right modal supports no longer
coincide. In this setting, $\mathrm{Re}(\lambda_\star)$ determines
dynamical persistence, whereas $\mathrm{Im}(\lambda_\star)$ describes
oscillatory modulation of the same branch. Topological breakup and
dynamical degradation can therefore occur at different stages of an
attack: a directed network may lose strong connectivity while still
retaining a nonzero relaxation channel in the projected decaying
sector. The branch-aware construction addresses this mismatch by
following the spectral carrier of the relaxation process rather than
stopping at a purely combinatorial threshold.

In the reciprocal limit, the same principle reduces to the weighted
Fiedler sector. The slow branch becomes real, the biorthogonal
structure collapses to an orthogonal eigenmode, and the robustness
metric becomes the weakest diffusive channel of the network. This
limit gives a geometric interpretation of the method: high-susceptibility
nodes are those whose removal most strongly compresses the slow
Fiedler mode on the survivor graph. The directed case is therefore not
a separate construction, but a non-Hermitian extension of the same
spectral-response principle.

The numerical results support this interpretation across both directed
and weighted-undirected networks. MS does not identify a universal
structural motif such as a hub or a bridge. Instead, it identifies the
local support of the slow relaxation branch. Depending on the network
ensemble, this support can appear as a hub-bridge in heterogeneous
graphs, a shortcut junction in small-world graphs, or a distributed
mixing center in random graphs. This explains why adaptive centrality
attacks can fail to reproduce the same ordering: they target structural
proxies, whereas MS targets the spectral mode that controls the
long-time relaxation response.

The comparison with FGIA-type baselines further clarifies the role of
the perturbation object. Spectral-perturbation rankings represented by
FGIA contain genuine modal information, but they are naturally derived
for fixed-dimensional edge modifications. Node removal is different:
it deletes a coordinate, removes all incident edges, and changes the
normalization of the survivor operator. The advantage of MS therefore
does not come from using spectral information in a generic sense, but
from matching the spectral response calculation to the
dimension-changing nature of vertex deletion.

From a physical viewpoint, the MS score can be interpreted as a
branch-level modal load. Nodes with large susceptibility carry a large
biorthogonal contribution to the slow relaxation channel, so their
removal produces a strong reduction of the branch-tracked decay rate.
Conversely, the same nodes are natural candidates for reinforcement
when the goal is to preserve the relaxation pathway. This suggests a
possible connection between functional attackability and spectral
control design, although explicit control or reinforcement strategies
are beyond the scope of the present work.

Several limitations define the range of validity of the framework. The
MS score is a first-order estimator of the branch response, and its
quantitative accuracy is expected to be highest when the selected slow
branch is spectrally isolated and not too strongly localized on a
single coordinate. In nearly degenerate spectra, strongly localized
modes, or regimes where node removal induces abrupt branch
reorganization, higher-order effects can reduce the accuracy of the
predicted drop. The exact one-step diagnostics show that the ranking
remains robust in the tested ensembles, but the method should still be
understood as a branch-aware susceptibility approximation rather than
as an exact greedy attack.

Overall, the main physical lesson is that functional vulnerability can
be carried by a spectral branch before it appears as the failure of a
structural component. A robustness theory for relaxation processes
should therefore be formulated at the level of dynamical branches, with
node importance defined by the contribution of each node to the
dominant decay channel.

\section{Conclusion}
\label{sec:conclusion}

We have introduced a modal-susceptibility framework for adaptive node attacks based on the slowest nonzero Laplacian branch. 
By defining robustness through $\mathrm{Re}(\lambda_2)$ and constructing a branch-tracked susceptibility score, the method provides a common spectral description of vulnerability in directed, undirected, and weighted undirected networks.

The key idea is to align the attack objective with the physical quantity that governs long-time dynamics. 
Rather than targeting structural proxies, MS ranks nodes according to their estimated impact on the decay channel that controls asymptotic relaxation. 
This leads to a consistent interpretation across symmetry classes: the directed case represents a non-Hermitian extension of the Fiedler-based picture in the undirected limit.

Beyond its algorithmic role, the framework highlights a broader principle: robustness, attackability, and controllability are naturally expressed in terms of spectral sensitivity of slow modes. 
This perspective provides a basis for analyzing transport, relaxation, and stability in asymmetric networks.

Future work will extend the approach to large sparse systems, incorporate ensemble-level validation, and explore applications in control and design. 
More generally, the present results suggest that spectral-response theory offers a natural language for studying dynamical robustness in complex networks.

\bibliographystyle{apsrev4-2}
\bibliography{prr_ms_directed_spectral_attack_refs}

\end{document}